\documentclass[dvips]{acta}
\usepackage{supertabular,lscape,epsfig}
\usepackage{amssymb}
\usepackage{amsmath}
\DeclareSymbolFont{ppa}{OT1}{ppl}{m}{it}
\DeclareMathSymbol{\vv}{\mathalpha}{ppa}{'166}

\newfont{\hb}{rphvb at 10pt}
\newfont{\hbo}{rphvbo at 10pt}
\newfont{\bitt}{rptmbi at 12pt}
\newfont{\bits}{rptmbi at 11pt}

\SetPages{65}{75}

\SetVol{64}{2014}

\begin{document}

\newcommand{\TabApp}[2]{\begin{center}\parbox[t]{#1}{\centerline{
  {\bf Appendix}}
  \vskip2mm
  \centerline{\small {\spaceskip 2pt plus 1pt minus 1pt T a b l e}
  \refstepcounter{table}\thetable}
  \vskip2mm
  \centerline{\footnotesize #2}}
  \vskip3mm
\end{center}}

\newcommand{\TabCapp}[2]{\begin{center}\parbox[t]{#1}{\centerline{
  \small {\spaceskip 2pt plus 1pt minus 1pt T a b l e}
  \refstepcounter{table}\thetable}
  \vskip2mm
  \centerline{\footnotesize #2}}
  \vskip3mm
\end{center}}

\newcommand{\TTabCap}[3]{\begin{center}\parbox[t]{#1}{\centerline{
  \small {\spaceskip 2pt plus 1pt minus 1pt T a b l e}
  \refstepcounter{table}\thetable}
  \vskip2mm
  \centerline{\footnotesize #2}
  \centerline{\footnotesize #3}}
  \vskip1mm
\end{center}}

\newcommand{\MakeTableApp}[4]{\begin{table}[p]\TabApp{#2}{#3}
  \begin{center} \TableFont \begin{tabular}{#1} #4 
  \end{tabular}\end{center}\end{table}}

\newcommand{\MakeTableSepp}[4]{\begin{table}[p]\TabCapp{#2}{#3}
  \begin{center} \TableFont \begin{tabular}{#1} #4 
  \end{tabular}\end{center}\end{table}}

\newcommand{\MakeTableee}[4]{\begin{table}[htb]\TabCapp{#2}{#3}
  \begin{center} \TableFont \begin{tabular}{#1} #4
  \end{tabular}\end{center}\end{table}}

\newcommand{\MakeTablee}[5]{\begin{table}[htb]\TTabCap{#2}{#3}{#4}
  \begin{center} \TableFont \begin{tabular}{#1} #5 
  \end{tabular}\end{center}\end{table}}

\newfont{\bb}{ptmbi8t at 12pt}
\newfont{\bbb}{cmbxti10}
\newfont{\bbbb}{cmbxti10 at 9pt}
\newcommand{\uprule}{\rule{0pt}{2.5ex}}
\newcommand{\douprule}{\rule[-2ex]{0pt}{4.5ex}}
\newcommand{\dorule}{\rule[-2ex]{0pt}{2ex}}
\begin{Titlepage}
\Title{Transiting Planets Orbiting Source Stars in Microlensing Events}
\vspace*{11pt}
\Author{K.~~R~y~b~i~c~k~i$^1$~~ and~~ £.~~W~y~r~z~y~k~o~w~s~k~i$^{1,2}$}
{$^1$Warsaw University Observatory, Al. Ujazdowskie 4, 00-478 Warszawa, Poland\\
e-mail: (krybicki,lw)@astrouw.edu.pl\\
$^2$Institute of Astronomy, University of Cambridge, Madingley Road, CB3 0HA Cambridge, UK}
\Received{December 20, 2013}
\end{Titlepage}

\vspace*{11pt}
\Abstract{The phenomenon of microlensing has successfully been used to detect 
extrasolar planets. By observing characteristic, rare deviations in the
gravitational microlensing light curve one can discover that a lens is a
star--planet system. In this paper we consider an opposite case where the
lens is a single star and the source has a transiting planetary
companion. We have studied the light curve of a source star with transiting
companion magnified during microlensing event. Our model shows that in
dense stellar fields, in which blending is significant, the light drop
generated by transits is greater near the maximum of microlensing, which
makes it easier to detect. We derive the probability for the detection of a
planetary transit in a microlensed source to be of $2\times10^{-6}$ for an
individual microlensing event.}{planetary systems -- Gravitational lensing:
micro}

\vspace*{11pt}
\Section{Introduction}
The search for extrasolar planets is a very dynamically developing branch
of modern astrophysics. After the discovery of the first planet (Wolszczan
and Frail 1992), and the detection of the first planet orbiting a
solar-type, main sequence star (Mayor and Queloz 1995) several planet
detecting methods have been developed. There are two natural phenomena,
which can be used for planetary detection, namely gravitational
microlensing and transits. In this work we consider the advantages of both
phenomena occurring together -- the case of gravitational microlensing of a
transiting planetary system (hereafter MiTr).

First planetary transit was detected in 1999 (Charbonneau \etal 2000), but
it was only a confirmation of the existence of a planet -- HD\,209458b had
already been detected by the means of the radial velocity
measurements. First planet ever discovered with the transit method was
OGLE-TR-56b (Udalski \etal 2002b, Konacki \etal 2003). Since then, the
transit method has been very successful and today it is one of the most
effective methods of discovering extrasolar planets.

The most problematic issue in planetary transits detection are the objects
mimicking such transits while being of completely different origin. Many
false-positives are observed in very crowded fields, in which blending with
neighboring stars is very common. If there is a star behind or in front of
an eclipsing binary star, eclipses are shallowed and hence the main eclipse
can look similar to a planetary transit (secondary eclipse is then lost in
the noise). Another example of false-positives are binaries generating
grazing eclipses (which can be as shallow as planetary transits) and binary
star systems in which one component is much smaller than the other. In this
work we consider the first one, the most common case of false-positives:
blended binary stars. If such object is microlensed, it is possible to
derive blending parameter and to answer the question whether the drop of
light during a transit is caused by a planet.

\hglue-4pt Gravitational microlensing phenomena can be directly used for exoplanets
search (Mao and Paczyñski 1991). When a lens consists of two components
(for example a star and a planet), it is possible to detect specific peaks
on regular microlensing light curves. Observing such phenomena can provide
the mass ratio of the components, which is crucial to determine whether it
is a planetary system or a binary star.  Microlensing phenomena are
incredibly rare, because almost perfect alignment of three objects has to
occur. Therefore observations must be conducted in very dense stellar
fields, most of all toward the Galactic bulge, to increase chances of
detecting microlensing events. For example, the OGLE-IV survey currently
detects about 2000 microlensing events every year, among which at least a
dozen shows planetary signatures (\eg Poleski \etal 2014).

In this paper we simulate the case, where a magnified source hosts a
planet. We consider a single point-like lens and derive the probability of
such configuration to happen and be detectable in currently on-going
microlensing surveys. First study of the microlensed planetary transit was
conducted by Lewis (2001), however they only considered an influence of
transits on the caustic crossing binary lens events.

The paper is organized as follows. In Section~2 we describe the model of
planetary transit and its microlensing as well as the model for the
accuracy of the photometry. Then in Section~3 we describe the results of
our simulations and derive the probability of the microlensed transit
source. We summarize the results in Section~4.

\Section{Model}
We consider a situation in which the light from a star (hence a source
star), which is being transited by a planet, is amplified due to
microlensing by a single lensing object.

\subsection{Planetary Transit}
Planetary transit is a straightforward geometrical problem, however, a few
simplifying approximations are needed. First of all, we assume that planet
moves along a straight line. In fact, its path is an ellipse, but compared
to the whole period, a transit usually lasts short enough for this
assumption to be reasonable. We also consider only circular orbits of
planets, because transit method is sensitive to planets orbiting very
close to their parent stars. Such configuration causes strong tidal
effects and hence, circularization of the orbit.

We then consider the following surface brightness $I(r)$ model, which
includes simple approximation of limb darkening (Heyrovsky 2007):
$$I(r)=I_0\left(1-\Gamma\left(1-\sqrt{\frac{R_{\star}^2-r^2}{R_\star^2}}\right)\right).\eqno(1)$$
Intensity of radiation {\it I} depends on the distance $r$ from the star
center, stellar radius $R_*$ and limb darkening coefficient $\Gamma$. The
latter is defined by the surface luminosity values on the center and on the
limb:
$$\Gamma=\frac{(I(0)-I(R_{\star}))}{I(0)}.\eqno(2)$$
Value of this coefficient varies from 0 (for the star disk equally luminous
from center to the edge) to 1 (for the star disk which surface brightness
is zero on the edge). We assume that stars and planets are perfectly
spherical -- we do not take into account the fact that rotating spherical
bodies are typically flattened on their poles.

It is obvious that the apparent luminosity drop during a transit depends on
the ratio of angular sizes of the planet and the star, which does not
depend on the distance to the system. Thus, it does not matter whether a
planet is orbiting 0.01~a.u., 5~a.u., or 10~a.u. from its parent star --
depth of transit will be always the same (although semi-major axis has an
indirect influence on transit's duration).

Our model of a transit has four main parameters: stellar radius $R_*$,
planetary radius $R_p$, orbital period $P$ and semi-major axis $a$. Apart
from those there are also other parameters like blending parameter $f$,
limb darkening coefficient $\Gamma$ and orbital inclination $i$. The flux
of radiation, when neglecting transits, is given by:
$$F_{\rm max}=F_1+F_{\rm bl}\eqno(3)$$
where $F_1$ is the flux of the star, while $F_{\rm bl}$ is the flux of
the third light (blend). Now we can define time-dependent flux of the
system during the transit:
$$F_{\rm tr}(t)=F_{\rm max}-\Delta F(t),\eqno(4)$$
$\Delta F(t)$ is the flux deficit caused by the passage of a planet in
front of the disk of the star. If surface brightness of the star $I_1$
were constant, we would have the following formula for the flux deficit:
$$\Delta F(t)=I_1\pi R_p^2\alpha(t)=F_1\left(\frac{R_p}{R_{\star}}\right)^2 
\alpha (t),\eqno(5)$$
$\alpha(t)$ is the ratio between the overlapping areas of the stellar and
planetary disks to the whole planetary disk area. Including the limb
darkening from (1) gives:
$$\Delta F(t)=\iint_{S(t)} I(r) \dd S.\eqno(6)$$
Integration is done over the surface $S(t)=\pi R_p^2\alpha(t)$. The
calculations of $\Delta F(t)$ are time consuming, since Monte Carlo method
is used in the surface integrals.

Now, we convert the depth of a transit to magnitudes, as a function of
time:
$$\Delta m_{\rm tr}(t)=m_{\rm tr}(t)-m_{\rm max}=-2.5\log
\left(\frac{F_{\rm tr}}{F_{\rm max}}\right)
=-2.5\log\left(1-f\frac{\Delta F(t)}{F_1}\right)\eqno(7)$$
where $f$ is the blending parameter, defined as
$$f=\frac{F_1}{F_1+F_{bl}}.\eqno(8)$$
Blending parameter varies from 0 to 1. For $f=1$ the contribution of
blending light to the flux is negligible, while $f\rightarrow0$ means that
the blend is dominating. Simulation of a transit is done by calculating
$\Delta m$ for many consecutive instants of time, corresponding to changing
positions of the planet.

Eventually, we obtain the depth of a transit as a function of time and seven
parameters ($R_*$, $R_p$, $P$, $a$, $f$, $\Gamma$, $i$). We can now create
synthetic light curves showing the apparent brightness decrease during the
transit. Fig.~1 shows examples of synthetic light curves for different
sets of initial parameters.
\begin{figure}[htb]
\includegraphics{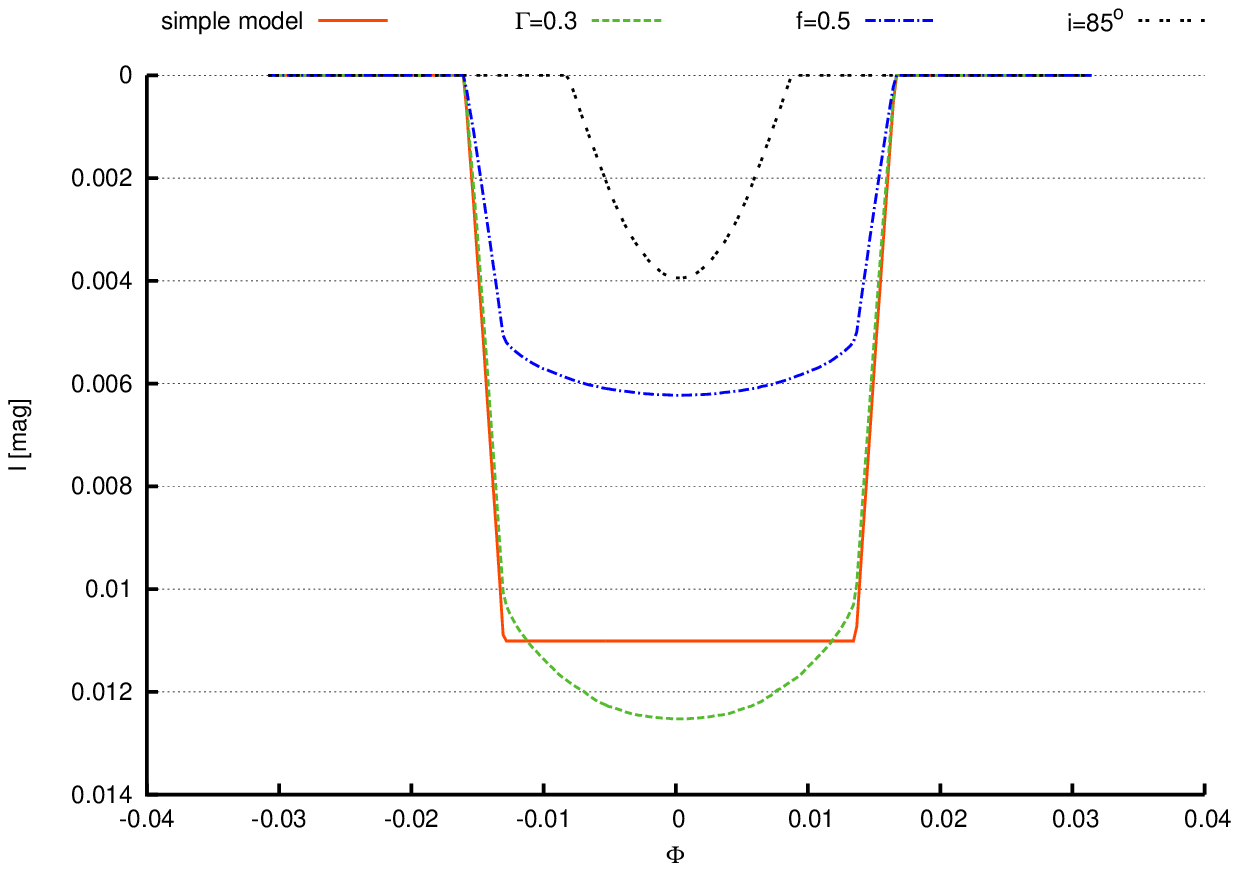}
\FigCap{Shape of a planetary transit. Red curve (solid) presents 
the simplest simulation: inclination $i=90\arcd$, without limb darkening
and blending. The green one (dashed) includes limb darkening, while the
blue curve (dash-dotted) shows the influence of blending. The black 
double-dotted line presents all the effects together for a case of
inclination of 85\arcd.}
\end{figure}
One can notice the influence of the blending parameter on the shape of the
light curve. For $f<1$ ($f=0.5$ here) the transit is shallowed, which is
crucial for our further considerations. If we observed a blended eclipsing
binary stars system, which generates light drop much greater than
planetary transit, such shallowing can make this binary star to mimic
star--planet system.

\subsection{Microlensing of a Transit}
We have derived the formula for the flux $F_{\rm tr}$ of the system
during planetary passage in front of the stellar disk (4), and its flux
$F_{\rm max}$ outside the transit (3). Let us now add microlensing to the
picture:
$$F_{\rm max}(t)=A(t)F_1+F_{\rm bl}=F_1\left(A(t)+\frac{1}{f}-1\right),\eqno(9)$$
$$F_{\rm tr}(t)=F_{\rm max}-A(t)\Delta F(t).\eqno(10)$$
Function $A(t)$ describes amplification due to gravitational microlensing. 

The amplification in the parametrization of Paczyñski (1996) is given by:
$$A(t)=\frac{u^2(t)+2}{u(t)\sqrt{u^2(t)+4}}\eqno(11)$$
where $u$ is the source-lens distance in units of the Einstein radius,
projected on the lens plane. The value of $u$ varies due to the relative
motion of the source and the lens, with $u=u_0$ at the closest approach of
these two objects. We can describe parameter $u$ using Einstein time $t_E$
and $u_0$:
$$u(t)=\sqrt{u_0^2+\left(\frac{t-t_0}{t_E}\right)^2}.\eqno(12)$$
Parameter $t_0$ denotes the moment of the highest amplification, when
$u=u_0$. We choose $t_0=0$ in the following analysis.

Microlensing alters the expression for the depth of transit:
$$\Delta m(t)=-2.5\log\left(\frac{F_{\rm max}(t)}{F_{\rm tr}(t)}\right)=
-2.5\log\left(\frac{A(t)F_1+F_{\rm bl}}{A(t)F_1+F_{\rm bl}-A(t)\Delta F(t)}\right).\eqno(13)$$

After rewriting $F_{\rm bl}$ using the blending parameter $f$ we obtain the
final formula for the depth of the transit:
$$\Delta m(t)=-2.5\log\left(\frac{A(t)+\frac{1}{f}-1}
{A(t)+\frac{1}{f}-1-A(t)\frac{\Delta F(t)}{F_1}}\right).\eqno(14)$$

Finally, the light curve including effects of microlensing and transits is
given by:
$$m(t)=m_{\rm max}+\Delta m(t)\eqno(15)$$
while for transits alone it is $m_{\rm tr}(t)=m_{\rm max}+\Delta m_{\rm
tr}(t)$ (\cf Eq.~7). Fig.~2 shows an example of a light curve
generated by our model.
\begin{figure}[htb]
\includegraphics{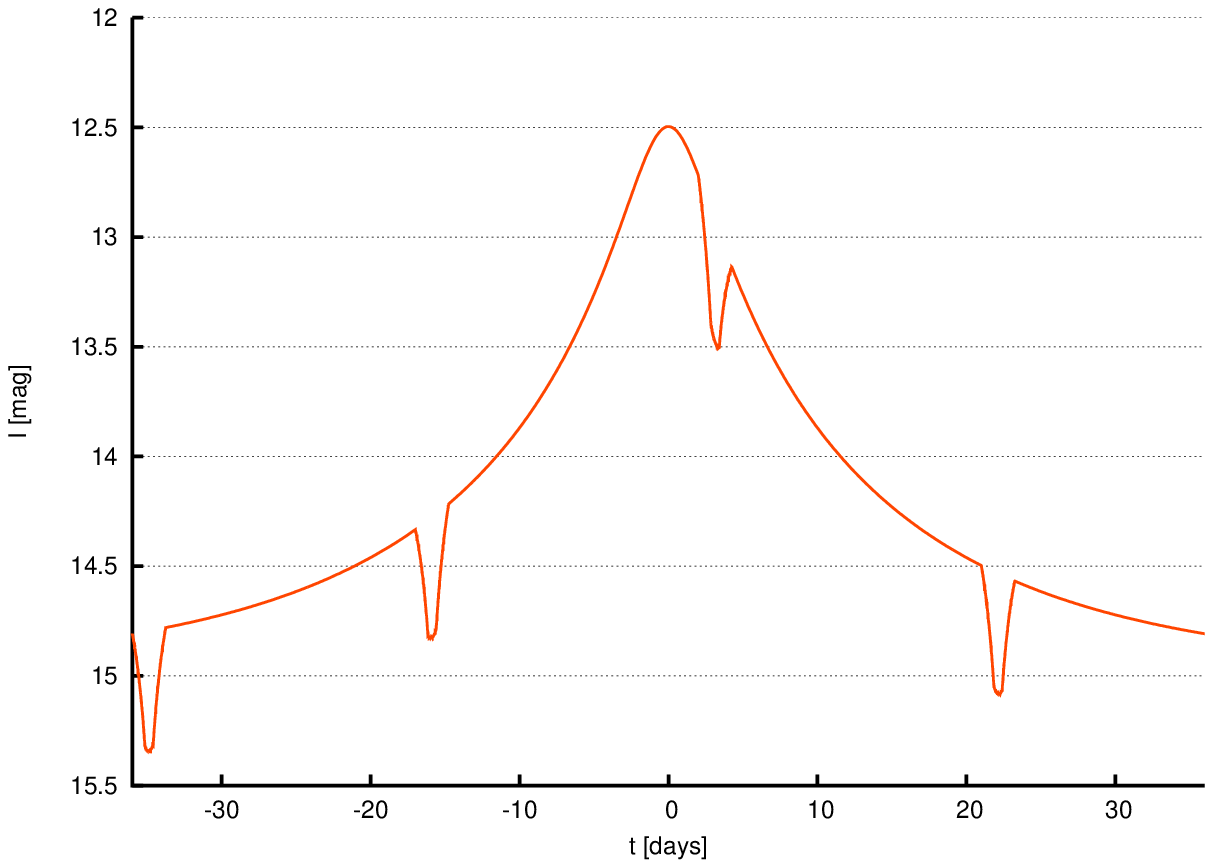}
\FigCap{Exemplary MiTr phenomenon. Light curve was generated for the 
following set of initial parameters: $u_0=0.1$, $f=1$, $\Gamma=0.3$,
$R_*=1~\RS$, $Rp=0.6~\RS$, $T=15$~d, $a=0.02$~a.u., $I_0=15$~mag,
$t_E=28$~d. This set was selected so that the characteristic shape of the
MiTr light curve was clearly seen on the plot. Therefore the drop of
brightness is much greater than typical caused by a planetary transit.}
\end{figure}

\subsection{Accuracy of the Photometry}
To derive constraints on the detection of microlensed transiting sources,
we need to know how precisely we can measure depth of transits. In this
work we assume that uncertainties of the photometry are similar to those in
the Galactic bulge fields data of the OGLE-III project (Udalski \etal
2002a). In that survey the typical uncertainty for a 15~mag source was
about 0.005~mag. In order to scale the error-bar with the brightness during
its change in a microlensing event we use the empirical formula from
Wyrzykowski (2005) (see also Wyrzykowski \etal 2009):
$$\Delta I=\Delta I_010^{0.33875(I-I_0)}\eqno(16)$$ 
where $\Delta I$ is the uncertainty of the brightness {\it I} and $I_0$ is
a normalizing brightness for which $\Delta I_0$ is known. Using the fact that
for $I_0=15$~mag $\Delta I_0=0.005$~mag, we can compute the error-bars for
any simulated magnitude. Top panel of Fig.~3 shows how $\Delta I$
changes in a microlensing event with maximum amplification at $t=0$.
\begin{figure}[htb]
\includegraphics{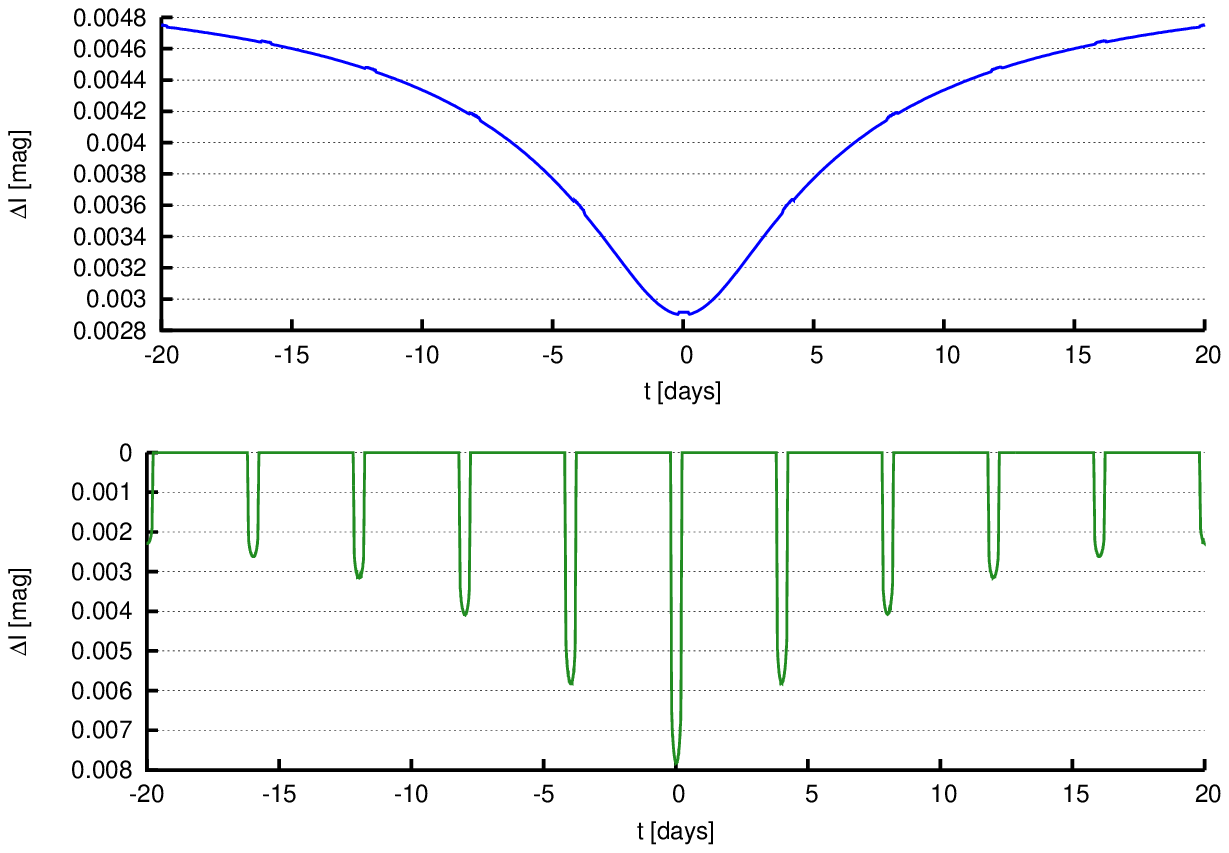}
\vskip4pt
\FigCap{{\it Top}: Changes of the uncertainty of photometric 
measurements during the MiTr phenomenon. Initial parameters describing
microlensing events were $u_0=0.1$, $t_0=0$~d, $t_E=28$~d,
$I_0=15$~mag. Significant blending has been added ($f=0.1$). {\it Bottom}:
Changes in transits depths during the microlensing event. This light curve
was generated for the same parameters as above.}
\end{figure}

In the case of a blended event with the source exhibiting variability in the
form of transits, the amplitude of that variability will increase with the
amplification (Wyrzykowski \etal 2006). In the limit of infinite
amplification, the amplitude of the variability reaches its completely
de-blended value. In other words, during the microlensing we can measure
the depth of the transit as if we used a much larger telescope with much
higher spatial resolution and no blending from nearby stars nor the
lens. In reality, the amplitude changes with amplification, depending on
the amount of blending. Bottom panel of Fig.~3 shows how the amplitude of
transit varies with amplification for maximum amplification of $A\approx
10$ ($u_0=0.1$) and blending parameter $f=0.1$. The change in amplitude
combined with the increase of accuracy of the photometry during a
microlensing event are the basis of our argument for a feasibility of the
detection of a planet transiting the source during microlensing event.

\section{Results}
\subsection{Simulations of MiTrs}
\begin{figure}[b]
\includegraphics[bb=45 75 400 270]{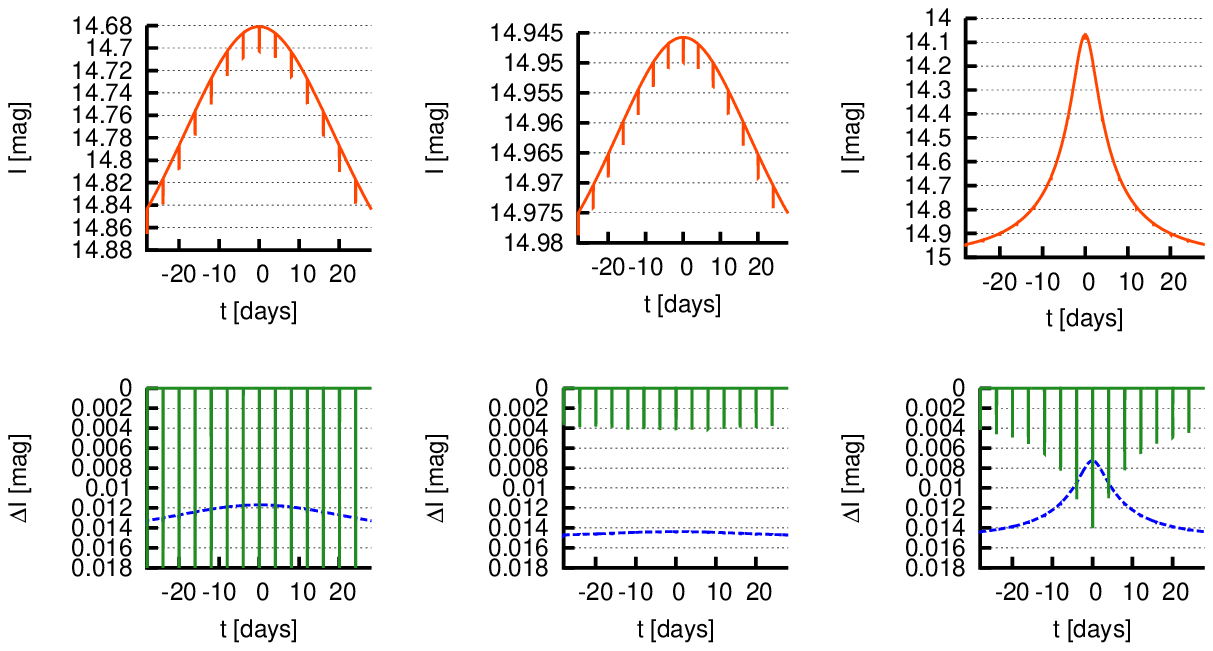}
\FigCap{Three simulations of interesting cases of MiTr phenomena for Hot Jupiter 
orbiting a small star. The following initial parameters were applied here:
$R_*=0.75~\RS$, $R_p=1~R_{\rm Jup}$, $T=4$~d, $a=0.05$~a.u., $I_0=15$~mag,
$t_E=28$~d, $i=90\arcd$. {\it Top panels} show synthetic light curves for
three different sets of blending and microlensing parameters. On the {\it
left} both microlensing magnification and blending are negligible ($u_0=1$,
$f=1$), in the {\it middle panels}, microlensing magnification is small
($u_0=1$), but blending is significant ($f=0.15$). {\it Right panel}
presents the most interesting case: both blending and microlensing
magnification are significant ($u_0=0.1$, $f=0.15$). {\it Bottom panels}
show depth of transits (green solid lines) and the level of three times
uncertainty of the photometry (blue dashed lines), calculated for the same
values of $f$ and $u_0$ as for light curve above. We are unable to detect a
transit in the case of a strong blending ({\it middle panels}), but if the
magnification is strong enough the transit can be detectable near the
maximum ({\it bottom right panel}). Note that there are different vertical
scales in the {\it upper panels}.}
\end{figure}
We simulated microlensed transits for a range of parameters of the
planetary systems and microlensing events. The values for microlensing
parameters were drawn in each simulation from the distributions obtained
for the 3500 standard microlensing events found in the OGLE-III data
(Wyrzykowski \etal in preparation), providing realistic statistics 
of parameters for microlensing events toward the Galactic bulge.

Fig.~4 shows examples of simulated microlensing events with planetary
transits for a selection of interesting combinations of their parameters.
The planet radius is arbitrarily set to $R_p=1~R_{\rm Jup}$. We show the
synthetic light curve of the event in each top panel. The comparison of the
depth of microlensed transits to the photometric uncertainty, $3\Delta I$,
is presented in the bottom panel. One can easily see when the transits
become detectable -- once their depth exceeds the typical error-bar of the
measurement (set as three standard deviations).

Left panels of Fig.~4 show the most common situation in sparse fields: in
case of very little blending (\ie when in the baseline light is composed
only of the source and its transits) the microlensing amplification only
shifts the brightness to a higher level of brightness, increasing slightly
the precision of the photometry. Middle panels of Fig.~4. show the case
when the source star with transits is severely blended and the transits are
shallowed to the limit beyond detectability. However, once microlensed, the
transits become significantly deeper due to the fact that the source star
becomes much brighter and starts to dominate over the blending objects
(right panels of Fig.~4).

\subsection{Probability of Detection in Microlensing Surveys}
It is obvious that microlensing of a planetary transit is an extremely rare
phenomenon. Here we estimate the probability $P_{\rm mitr}$ for a detection
of the microlensed source with planetary transits among all detected
microlensing events. Again, we only consider here observations toward the
Galactic bulge (and in particular the Baade's Window), because only such
dense fields provide a significant number of microlensing events.

First component of the overall probability is the probability that a source
hosts a transiting planet ($P_{\rm tr}$). For Hot Jupiters (HJs) considered
here we assumed $P_{\rm tr}=1/310$ as calculated by Gould \etal (2006),
based on the observational data. This is the probability derived for
Galactic disk stars, while our simulations are performed for the Galactic
bulge. The rate of transiting HJs for the central part of our Galaxy is
probably different than calculated by Gould \etal (2006). Even though, we
use those calculations as a reasonable approximation of $P_{\rm tr}$ in the
Galactic bulge.

Second component is $P_{\rm det}$ -- a probability that a MiTr event will
have at least two detectable transits. We used our model of the MiTr to
derive this probability in the following way. We drew the brightness in the
baseline $I_0$, blending parameter $f$, impact parameter $u_0$ and the
event time scale $t_E$ from the observational distribution of microlensing
events as found in the OGLE-III data (Wyrzykowski \etal in preparation).
Then, from the Besançon model of the Galaxy (Robin \etal 2004), we obtained
the relation between the radius of the star $R_*$ and its brightness $I_s$,
where $I_s$ is the source brightness in the microlensing event derived from
$I_0$ and $f$. We only selected stars belonging to the dominant bulge
population, to assure we probe the most likely population of the source
stars. Combining all those parameters allowed us to simulate a MiTr event.
The simulation was performed with fixed period $P=4$~d, $a=0.05$~a.u. and
$R_p=1~R_{\rm Jup}$, as common parameters for HJ planet population. For
each set of parameters, probability was calculated for different phases of
transits and averaged. Using the MiTr model we generated 10\,000
microlensing events with varying microlensing and stellar radii
parameters. In each event we computed the depth of the transit for the
maximum amplification and compared it to the expected photometric
error-bars ($3\Delta I$), thus obtaining a probability of detection of each
case.

Last point to take into account is the fact that in a blended
microlensing event in the Galactic bulge there are on average 2.3 stars
which could potentially be microlensed in a typical OGLE-like resolution
image and with OGLE-like sensitivity (Wyrzykowski 2005, Koz³owski 2007,
Koz³owski private communication 2013). Every time we simulate a
microlensing of an object which consists of a few blended stars, we
assume, that microlensing is related to the one with a transiting
companion. In reality, each component of blended light source has the
same probability to be magnified by means of microlensing. Thus the
$P_{\rm bl}=1/2.3$ factor has to be included in the overall probability.
We need to account for all those sources with detectable transits
within the blend which do not end up to be microlensed.

The final probability for occurrence of a detectable planetary transit of
HJ in a microlensing event is given by the product $P_{\rm mitr}=P_{\rm
det}\times P_{\rm tr}\times P_{\rm bl}$. For HJs, considered in our
simulations, we obtain the probability of $P_{\rm mitr}=
2\cdot10^{-6}$. Hence, we should detect approximately two MiTr of HJs among
every million microlensing events. With the current rate of microlensing
events detections of about 2000 per year (OGLE), it is very unlikely that
MiTr event will be found soon. OGLE survey so far, during all its phases,
have found about 10\,000 microlensing events (with nearly 6000 during
OGLE-IV alone), hence there is probably no microlensed transit present in
the archival data. Nevertheless, a possibility of seeing MiTr in the well
sampled future microlensing events, should be considered when, for example,
analyzing an anomaly at the top of a microlensing event.

\Section{Summary and Conclusions}
In this paper we presented theoretical predictions about the microlensing
of the source hosting a transiting planet. One of the main properties of
this novel method for finding planets is a capability of breaking the
common degeneracy in the planetary radius estimate due to the third light
in eclipsing binaries. We show that microlensing in crowded fields can
bring up the transits which otherwise remain buried in the photometric
noise thanks to two combined effects: increase in brightness, hence
improvement in measurement accuracy, and increase in the transit depth due
to de-blending of the lensed source. We estimated the probability of such a
phenomenon to $2\cdot10^{-6}$ in a survey with properties similar to the
OGLE project. This result means that most likely such an interesting
phenomenon has not yet been observed, and will not be observed in the near
future. Though our simulations yield somewhat negative result, information
that MiTr signal is most likely not present in the data could be useful,
\ie for eliminating potential sources of unknown irregularities in light
curves of microlensing events.

\Acknow{The authors would like to thank Profs. Andrzej Udalski and
Micha{\l} Jaroszy{\'n}ski for their comments and suggestions, which
significantly improved this work. This work was partially supported by
the Polish National Science Center (NCN) under the Grant No
2012/06/M/ST9/00172.}

\end{document}